\documentclass[10pt,twocolumn,aps,pra]{revtex4-1} %% For arXiv

\usepackage{amsthm}
\usepackage{amsmath}            %serve per le subequazioni
\usepackage{amssymb}            %serve per il simbolo "marchio registrato", \circledR
\usepackage{mathtools}
\usepackage{graphicx}           %serve per le figure eps, ps etcyy
\graphicspath{ {./images/} }
\usepackage[caption=false]{subfig}
\usepackage{multirow}

\usepackage{xcolor}
\usepackage{bm}

\usepackage{tcolorbox}
\usepackage{enumitem}

\usepackage[colorlinks,citecolor=blue,urlcolor=blue,bookmarks=false,hypertexnames=true]{hyperref} 
\newcommand{\ket}[1]{| #1 \rangle}
\newcommand{\bra}[1]{\langle #1 |}
\newcommand{\proj}[1]{\ket{#1}\bra{#1}}
\newcommand{\ts}{\otimes}
\newcommand{\id}{\openone}
\newcommand{\tr}{\text{Tr}}
\newcommand{\A}{\text{A}}
\newcommand{\B}{\text{B}}
\newcommand{\assemb}[1]{\{#1\}}

\usepackage{ulem}

\newcommand{\comt}[1]{}
\newcommand{\sect}[1]{Sec.~#1}
\newcommand{\apx}[1]{Appendix~#1}
\newcommand{\eq}[1]{Eq.~(#1)}
\newcommand{\fig}[1]{Fig.~#1}
\newcommand{\tab}[1]{Table.~#1}

\begin{document}
\title{Benchmarking Quantum State Transfer on Quantum Devices \\ using Spatio-Temporal Steering}

\author{Yi-Te Huang}
\affiliation{Department of Physics and Center for Quantum Frontiers of Research \& Technology (QFort), National Cheng Kung University, Tainan 701, Taiwan}

\author{Jhen-Dong Lin}
\email{jhendonglin@gmail.com}
\affiliation{Department of Physics and Center for Quantum Frontiers of Research \& Technology (QFort), National Cheng Kung University, Tainan 701, Taiwan}

\author{Huan-Yu Ku}
\email{huan\_yu@phys.ncku.edu.tw}
\affiliation{Department of Physics and Center for Quantum Frontiers of Research \& Technology (QFort), National Cheng Kung University, Tainan 701, Taiwan}
%\affiliation{Theoretical Quantum Physics Laboratory, RIKEN Cluster for Pioneering Research, Wako-shi, Saitama 351-0198, Japan}

\author{Yueh-Nan Chen}
\email{yuehnan@mail.ncku.edu.tw}
\affiliation{Department of Physics and Center for Quantum Frontiers of Research \& Technology (QFort), National Cheng Kung University, Tainan 701, Taiwan}
%\affiliation{Theoretical Quantum Physics Laboratory, RIKEN Cluster for Pioneering Research, Wako-shi, Saitama 351-0198, Japan}

%\author{Neill Lambert}
%\affiliation{Theoretical Quantum Physics Laboratory, RIKEN Cluster for Pioneering Research, %Wako-shi, Saitama 351-0198, Japan}

%\author{Franco Nori}
%\affiliation{Theoretical Quantum Physics Laboratory, RIKEN Cluster for Pioneering Research, %Wako-shi, Saitama 351-0198, Japan}
%\affiliation{Department of Physics, The University of Michigan, Ann Arbor, 48109-1040 Michigan, %USA}

\begin{abstract}
    Quantum state transfer (QST) provides a method to send arbitrary quantum states from one system to another. Such a concept is crucial for transmitting quantum information into the quantum memory, quantum processor, and quantum network. The standard benchmark of QST is the average fidelity between the prepared and received states. In this work, we provide a new benchmark which reveals the nonclassicality of QST based on spatiotemporal steering (STS). More specifically, we show that the local-hidden-state (LHS) model in STS can be viewed as the classical strategy of state transfer. Therefore, we can quantify the nonclassicality of QST process by measuring the spatiotemporal steerability.
    We then apply the spatiotemporal steerability measurement technique to benchmark quantum devices including the IBM quantum experience and QuTech quantum inspire under QST tasks. The experimental results show that the spatiotemporal steerability decreases as the circuit depth increases, and the reduction agrees with the noise model, which refers to the accumulation of errors during the QST process. Moreover, we provide a quantity to estimate the signaling effect which could result from gate errors or intrinsic non-Markovian effect of the devices.
\end{abstract}

\maketitle

\section{Introduction}\label{sec:intro}

%Quantum mechanics promises more powerful computation than classical ones by employing superposition and entanglement. IBM has developed some prototypes of superconducting 1-, 5-, 16-, 20-, 28-, and 50-qubits and are accessible through their web-interface or qiskit (a python package). Many algorithms have been proven mathematically faster than the best known classical algorithms and can be executed on IBM's devices.
%Quantum advantage relies on the properties of the superposition and entanglement and can take advantage on the classical 

%Quantum computation outperforms the classical computation due to the fundamental properties of the quantum mechanical, e.g. superposition and entanglement. Recently, noisy intermediate-scale quantum (NISQ) devices have been widely implied in quantum computations. IBM has developed some prototypes of superconducting 1-, 5-, 16-, 20-, and 50-qubits and are accessible through their web-interface or qiskit (a python package). There are many algorithms have proven mathematically faster than the best known classical algorithms and can be executed on IBM's devices. These algorithms critically rely on the demand of quantum computer, such as qubit coherent time, operation fidelity, and measure accuracy. We have to make sure that quantum computers are ready to run these algorithms, since the computational process suffers great noise. Therefore, benchmarks for applications (or algorithms) metrics are important to evaluate the efficiency and accuracy of quantum computation.

A reliable quantum state transfer (QST) from the sender to receiver is an important protocol for both quantum communication and scalable quantum computation~\cite{Kay2010,Christandl2004}. Such a process can not only be used to transmit quantum information between two computational components~\cite{Lvovsky2009,Yuan2019,Rosset2018}, but also to change the entanglement distribution in the quantum internet~\cite{Cirac1997,Chiribella2009,Hahn2019, Khatri2021}. To implement the QST, one can rely on the SWAP operation~\cite{Lu2019} or the quantum teleportation~\cite{Bennett1993} between the sender and the receiver. For hybrid quantum systems, e.g., phonons in ion traps~\cite{SchmidtKaler2003}, spin chain~\cite{Yao2011,Kandel2019,Lorenzo2013,Bayat2010,Bayat2014}, electro-optic~\cite{Rueda2019}, circuit quantum electrodynamics~\cite{Liu2017}, and bosonic quantum systems~\cite{Lau2019}, interaction between the sender and receiver through the communication line is required.

A similar concept is known as spatial steering, which states that the quantum states can be remotely prepared using entangled pairs. It was first proposed by Schr\"odinger~\cite{Schrodinger1935} against the famous thought experiment called the Einstein-Podolsky-Rosen paradox~\cite{Einstein1935}. The mathematical formulation of spatial steering was proposed recently~\cite{Wiseman2007,Uola2020,Cavalcanti2009,Cavalcanti2016}. Spatial steering plays a crucial role in many quantum information tasks, such as the channel discrimination problem~\cite{Piani2015,Sun2018, Ku2020},
one-sided quantum key distribution~\cite{Branciard2012}, measurement incompatibility~\cite{Cavalcanti2016,Uola2014,Quintino2014,Shin-Liang2016-2,Uola2015}, and no-cloning principle~\cite{Chiu2016}. Similar to the analogy between Bell and Leggett-Garg (LG) inequalities~\cite{Bell1964,Brunner2014,Leggett1985,Emary2014}, temporal steering~\cite{YuehNan2014,Che-Ming2015,Bartkiewicz2016} is also proposed as the temporal analog of spatial steering. Such a nonclassical temporal quantum correlation can be used to quantify the non-Markovianity~\cite{Shin-Liang2016,Ku2016}, witness quantum scrambling~\cite{Lin2020}, and certify quantum key distribution~\cite{Bartkiewicz2016}. Recently, spatiotemporal steering (STS), which is defined similarly to the Bell-LG inequality~\cite{White2016}, was proposed~\cite{Shin-Liang2017} to certify the nonclassical correlations in a quantum network~\cite{Krivchy2020}. We also highlight that a steering task is said to be unsteerable when the steering resources can be described by the local-hidden-state (LHS) model~\cite{Rodrigo2015,Uola2018}.

In this work, we define the classical strategy of state transfer process. In general, such a strategy can be mathematically described by the LHS model. Therefore, we employ the quantification of spatiotemporal steerability to quantify the nonclassicality of the QST process. Note that similar discussions regarding the nonclassicality of quantum teleportation protocol have been proposed in Refs.~\cite{Cavalcanti2017,Carvacho2018,ifmmode2019}.
%Moreover, we show the advantages of employing the spatiotemporal steerability.
%constructed by an ensemble of ontic states together with a stochastic map.

We then utilize the quantification of spatiotemporal steerability (or the QST nonclassicality) to benchmark noisy intermediate-scale quantum devices~\cite{Preskill2018} including the IBM quantum experience~\cite{IBMQ} and QuTech quantum inspire~\cite{Qutech}. Such quantum devices can now be applied to implement some quantum algorithms~\cite{Devitt2016,Kandala2017,Steiger2018,Knill2005} and simulations~\cite{Smith2019,GarcaPrez2020}.
In general, benchmarks of quantum devices provide us with a simple method to evaluate the performance of the quantum devices under certain quantum information tasks, e.g., benchmarking the shallow quantum circuits~\cite{Benedetti2019}, nonclassicality for qubit arrays~\cite{Waegell2019}, quantum chemistry~\cite{McCaskey2019}, and quantum devices~\cite{Klco2019,Wright2019,Bai2018}.
Our experimental results show that the degree of QST nonclassicality decreases as the circuit depth increases. In addition, the decrease agrees with the noise model, which describes the accumulation of noise (qubit relaxation, gate error, and readout error) during the QST process. In general, the results for the IBM quantum experience show that it outperforms QuTech quantum inspire from the viewpoint of QST nonclassicality. In addition, the results from IBM quantum experience are obtained before and after the maintenance. The result before the maintenance violates the no-signaling in time condition~\cite{Halliwell2016,Kofler2013,Knee2016,CheMing2012,Uola2019}, which is possibly due to the gate error and the non-Markovian effect in the devices~\cite{Morris2019,Pokharel2018,Ku2020-2}.

\section{Benchmarking Quantum State Transfer with spatiotemporal Steering}\label{sec:Benchmark}

In this section, let us briefly recall the quantum state transfer (QST) and the spatiotemporal steering (STS) scenario~\cite{Shin-Liang2017} in terms of the language of quantum information science. We will also discuss the similarities between them and demonstrate how to quantify the nonclassicality of QST process in the context of STS.

\subsection{Quantum state transfer}\label{sec:qst}

The protocol for QST is depicted by a sender (Alice) who prepares an arbitrary quantum state $\rho_{\A_0}$ and a receiver (Bob) who then receives the transferred state $\rho_{\B}$. Without loss of generality, the state transfer process can be described using a global quantum channel $\Lambda_t$, such that the prepared state $\rho_{\A_0}$ and the received state $\rho_{\B}$ are related based on the following equation:
\begin{equation}\label{eq:QST_rho_B}
    \rho_{\B}=\tr_{\A}\left[\Lambda_{t}(\rho_{\A_0} \otimes \sigma_{\B_0})\right],
\end{equation}
where $\sigma_{\B_0}$ is the initial state of Bob. Here, we use the  subscript $t$ to represent the time which will be used later. The QST process $\Lambda_t$ is perfect if $\rho_{\A_0}$ and $\rho_{\B}$ are related by an unitary operation, of which the inverse, i.e., the decoding unitary, can be used to recover the prepared state $\rho_{\A_0}$~\cite{Kay2010}. We note that \eq{\ref{eq:QST_rho_B}} can be easily applied to $d$-level~\cite{Bayat2014, Liu2017} and multi-partite systems~\cite{Lorenzo2013}.

For qubit systems, the states can be perfectly transferred in a spin chain model with $XY$ coupling~\cite{Yao2011,Christandl2004, Lorenzo2013}, $XYZ$ coupling~\cite{Kandel2019}, or $XXZ$ coupling~\cite{Bayat2010}. We will experimentally present an explicit example in the cloud based on the $XY$ interaction to implement the QST process in \sect{\ref{sec:quantum state transfer process}}.
%With a suitable unitary operation, the quantum state hold on Bob can be recovered to the one which Alice sends with the prior knowledgment of the quantum channel $\Lambda_t$.
%Given a quantum channel $\Lambda$ which is capable of performing quantum state transfer (QST), the protocol for the state transfer task can be described in the following three steps: First, Alice prepares an arbitrary state $\ket{\psi}$ (or density matrix $\rho_\psi$) that she wants to transfer to Bob. Second, Alice send the prepared state through a quantum channel $\Lambda(\rho_\psi)$ and Bob receives the state. Finally, before Bob extracts the information of the system, he performs some operations to "fix up" the state based on the knowledge of the channel $\Lambda$.

\subsection{Spatiotemporal steering}\label{sec:STS}

In the STS scenario, a bipartite system is shared by Alice and Bob. At initial time $t=0$, Alice performs local measurements labeled as $x$ with the corresponding outcomes labeled as $a$. After Alice's measurement, the bipartite system is then sent into a global quantum channel  $\Lambda_t$. Finally, Bob receives a state from a probability distribution over the set $\{\tilde{\varrho}_{a|x}(t)\}_{a,x}$. Without loss of generality, one can use the terminology in the standard spatial steering~\cite{Wiseman2007,Uola2020} which is termed as the assemblage $\{\varrho_{a|x}(t) \coloneqq P_{\A}(a|x)\tilde{\varrho}_{a|x}(t)\}_{a,x}$ to characterize the spatiotemporal steerability. Here, $P_{\A}(a|x)$ describes the probability of obtaining the output $a$ conditioned on Alice's choice of measurement $x$, and $\tilde{\varrho}_{a|x}(t) = \varrho_{a|x}(t)/P_\A(a|x)$ is the conditional quantum state received by Bob. According to quantum theory, when Alice and Bob share an initial separable state, i.e., $\sigma_{\A_0}\ts\sigma_{\B_0}$, the states received by Bob after the channel can be expressed as
\begin{equation}\label{eq:STS_rho_B}
\tilde{\varrho}_{a|x}(t)=\tr_{\A}\left[\Lambda_{t}(\tilde{\varrho}_{a|x}(0) \ts \sigma_{\B_0})\right],
\end{equation}
where $\tilde{\varrho}_{a|x}(0)=(M_{a|x} \sigma_{\A_0} M^\dagger_{a|x})/P_\A(a|x)$, $\sigma_{\A_0}$($\sigma_{\B_0}$) is the initial state for Alice (Bob), and $\assemb{M_{a|x}}$ is considered to be a set of projective measurements. Throughout this work, we consider $\sigma_{\A_0}$ to be $\id/d$ for satisfying no-signaling in time condition in \eq{\ref{eq:no-signaling}}, which we will explicitly discuss later.

We call the assemblage $\assemb{\varrho_{a|x}(t)}_{a,x}$ spatiotemporal unsteerable if it agrees with the local-hidden-state (LHS) model~\cite{Jones2007,Wiseman2007}, namely
\begin{equation}\label{eq:LHS}
    \varrho^{\text{LHS}}_{a|x}(t)=\sum_\lambda P(\lambda) P_{\A}(a|x, \lambda) \sigma(\lambda) ~~\forall~~ a, x,
\end{equation} such that the assemblage can be constructed by an ensemble of ontic states $\{P(\lambda),\sigma(\lambda)\}_\lambda$ together with the stochastic map $\{P_\A(a|x,\lambda)\}_\lambda$, which maps the local hidden variable $\lambda$ to $a|x$. In other words, an assemblage can be described by the LHS model, whenever it can be explained classically. Because the set of LHS models forms a convex set, we can, in general, quantify the spatiotemporal steerability by the notion of the spatiotemporal steering robustness $\mathcal{STSR}$~\cite{Shin-Liang2017, Ku2018-2}, which is defined as follows:
\begin{equation}\label{eq:robustness}
    \begin{split}
        &\mathcal{STSR}(\assemb{\varrho_{a|x}(t)}) = \displaystyle \min_{r, \assemb{\tau_{a|x}}, \assemb{\varrho^{\text{LHS}}_{a|x}(t)}} ~~r,\\
        &\text{s.t.} ~~~\frac{1}{1+r}\varrho_{a|x}(t)+\frac{r}{1+r}\tau_{a|x}=\varrho^{\text{LHS}}_{a|x}(t)~\forall~ a, x.
    \end{split}
\end{equation}
The optimal solution $r^*$ in \eq{\ref{eq:robustness}} can be interpreted as the minimal amount of noisy assemblage $\assemb{\tau_{a|x}}$ required to destroy the spatiotemporal  steerability of the underlying assemblage $\assemb{\varrho_{a|x}(t)}$.
The optimization problem can be computed by the semidefinite program presented in \apx{\ref{sec:SDP}}.

To obtain the spatiotemporal steerability quantum mechanically~\cite{Ku2018-2,Shin-Liang2017}, the assemblage should satisfy the no-signaling in time (NSIT) condition~\cite{Kofler2013,Clemente2015,CheMing2012}, that is, the underlying assemblage obeys the following condition:
\begin{equation}\label{eq:no-signaling}
    \sum_a \varrho_{a|x}(t) = \sum_a \varrho_{a|x'}(t) ~~~ \forall ~~~ x \neq x'.
\end{equation}
Once the NSIT condition is violated, the obtained spatiotemporal steerability can be explained by classical signaling effect. Actually, one can always violate the spatiotemporal steering inequality using additional classical communication from Alice to Bob. A similar situation has been reported as a communication loophole in the spatial steering scenario~\cite{Nagy2016} and the clumsiness loophole in the spatiotemporal/temporal quantum correlations~\cite{Uola2019,Ku2020-2}. 

Here, we provide a quantity $\mathcal{D}$ to estimate the signaling effect by using the trace distance
\begin{equation}\label{eq:signaling}
  \mathcal{D}(\assemb{\varrho_{a|x}(t)}) =\max_x~ \frac{1}{2}\left|\left|\sum_a \varrho_{a|x}-\sum_a \varrho_{a|x'}\right|\right|_1 ~\forall~ x \neq x'.
\end{equation}
By the definition in \eq{\ref{eq:no-signaling}}, the value of $\mathcal{D}$ is zero if and only if the given assemblage satisfies NSIT condition. Furthermore, we prove that $\mathcal{D}$ is a lower bound of $\mathcal{STSR}$, namely
\begin{equation}\label{eq:stsr_geq_D}
    \mathcal{STSR}(\{\varrho_{a|x}(t)\}) \geq \mathcal{D}(\{\varrho_{a|x}(t)\})
\end{equation}
(see \apx{\ref{sec:proof_quantify_NSIT}} for the derivation). Such a relation justifies that when $\mathcal{STSR} = \mathcal{D}$, the observed $\mathcal{STSR}$ can be alternatively falsified due to extra classical communication resource.

\subsection{Quantifying the nonclassicality of QST using STS}\label{sec:nonclassicality of QST}

We can observe that Bob's received states for STS and QST processes are basically in the same form [see \eq{\ref{eq:QST_rho_B}} and \eq{\ref{eq:STS_rho_B}}], indicating that a QST process can be discussed from the viewpoint of STS.
In fact, the classical strategy of state transfer can be defined as: Bob constructs the received state by an ensemble of ontic states together with a stochastic map. In other words, the LHS model in \eq{\ref{eq:LHS}} also describes the classical strategy of state transfer. Based on such insights, the $\mathcal{STSR}$ can also be used to quantify the nonclassicality of QST. 

Here, we provide a vivid example of the classical state transfer known as measure-and-prepare scenario~\cite{Horodecki2003}. Let us consider Alice is going to tranfer a set of quantum states $\assemb{\tilde{\rho}^\A_{a|x}}$ [sampled from a probability distribution $P_\A(a|x)$] to Bob. Alice first performs the measurement $M_\xi$ on her state $\tilde{\rho}^\A_{a|x}$ and obtain the outcome $\xi$ according to the distribution $P(a,\xi|x)=P_\A(a|x)\tr[M_\xi\tilde{\rho}^\A_{a|x}]$. After receiving $\xi$ from Alice, Bob can then construct the unnormalized states $\rho^\B_{a|x}$ by preparing a set of states $\assemb{\sigma_\xi}$ together with distribution $P(a,\xi|x)$, namely
\begin{equation}
    \rho^\B_{a|x} = \sum_\xi P(a,\xi|x) \sigma_\xi = \sum_\xi P(\xi) P(a|x, \xi) \sigma_\xi. \notag
\end{equation}
The above equation is mathematically equivalent to \eq{\ref{eq:LHS}}. Therefore, the measure-and-prepare scenario can be described by LHS model. 

We now point out that the $\mathcal{STSR}$ of the underlying assemblage is invariant under an arbitrary unitary transformation $\tilde{U}$, namely
\begin{equation}\label{eq:PST_SR}
  \mathcal{STSR}(\assemb{\varrho_{a|x}}) = \mathcal{STSR}(\assemb{\tilde{U}\varrho_{a|x}\tilde{U}^\dagger})
\end{equation}(see \apx{\ref{sec:proof_invariant}} for the derivation). Accordingly, the optimal decoding unitary, which is used to recover the prepared state, is redundant under STS scenario (see the discussion in \sect{\ref{sec:qst}}). In fact, neither the complete knowledge of the decoding unitary nor the full description of $\Lambda_t$ is required to quantify the nonclassicality of QST process. We recall that steering-type scenarios are one-sided device independent, in which Alice’s measurement devices cannot be characterized (untrusted)~\cite{Uola2020, Cavalcanti2016}. Note that the standard benchmark of a QST process, e.g., average fidelity, relies on the complete knowledge of the decoding unitary or, equivalently, $\Lambda_t$. In addition, to reconstruct $\Lambda_t$ and the decoding unitary, the quantum process tomography, which requires fully trusted devices, is necessary.

% The advantage of employing the $\mathcal{STSR}$ is twofold. First, we can show that the $\mathcal{STSR}$ of the underlying assemblage is invariant under an arbitrary unitary transformation $\tilde{U}$; i.e., 
% Accordingly, the local unitary operation for verifying the fidelity after the QST process is unnecessary when considering the STS scenario. Second, because steering-type scenarios are one-sided device independent~\cite{Uola2020, Cavalcanti2016}, it means that to certify the QST non-classicality from the perspective of STS, we do not have to characterize Alice's measurement operators $\assemb{M_{a|x}}$ and the post-measurement states $\tilde{\varrho}_{a|x}$  before sending the system into global operation $\Lambda_t$. Instead, Alice's measurement results are only summarized in the probability distribution $P_\A(a|x)$, and the transferred states for Bob are unknown to him before the state tomography is performed. Therefore, to certify the QST non-classicality, we do not have to compute the average fidelity, which needs to verify all possible prepared states. 

The experiemental setup for quantifying the nonclassicality of QST process using STS can be summarized in the following algorithm box:
\begin{tcolorbox}[title=Steerability Estimation]
  With the assemblage defined as $\{\varrho_{a|x}(t) \coloneqq P_{\A}(a|x)\tilde{\varrho}_{a|x}(t)\}$, the nonclassicality of a QST process $\Lambda_t$ can be quantified by the following steps:

  \begin{enumerate}[label=(\arabic*)]
    \item Alice performs projective measurements on maximally mixed states and get a probability distribution $P_\A(a|x)$.
    
    \item Alice apply $\Lambda_t$ on all post-measurement states $\assemb{\tilde{\varrho}_{a|x}(0)}$.
    
    \item Bob performs quantum state tomography to characterize a set of receiving quantum states $\assemb{\tilde{\varrho}_{a|x}(t)}$.
    
    \item Calculate $\mathcal{STSR}(\assemb{\varrho_{a|x}(t)})$.
  \end{enumerate}

  $\mathcal{STSR}(\assemb{\varrho_{a|x}(t)}) > 0$ certifies that $\Lambda_t$ is a nonclassical QST process.
  %$\mathcal{STSR}(\assemb{\varrho_{a|x}(t)}) > 0$ and $\mathcal{D}(\assemb{\varrho_{a|x}(t)}) = 0$ certifies that $\Lambda_t$ is non-classical. Note that $\Lambda_t$ is called a perfect quantum state transfer if and only if $\mathcal{STSR}(\assemb{\varrho_{a|x}(t)}) = \mathcal{STSR}(\assemb{\varrho_{a|x}(0)})$.
\end{tcolorbox}

%Therefore, due to the invariant of the $\mathcal{STSR}$ under the perfect QST and the property of the $\mathcal{STSR}$, which is the monotone under the CPTP map, we can quantify the process of the QST by considering the STS scenario. Here, we mean that the $\mathcal{STSR}$ can be used demonstrate the nonclassicality of QST process. In the classical state transfer process, which in general can be described by the measurement-and-prepare channel~\cite{}, the value of the $\mathcal{STSR}$ is always $0$. We note that if we use the fidelity as a metric to quantify the process of the QST, the value of the fidelity is not equal to $1$ even under the classical state transfer process.

%\thm{\ref{eq:PST_SR}} not only shows the perfect QST cannot change the value of the $\mathcal{STSR}$ but also the unnecessary recovered unitary operation $S$ due to the invariant under the unitary. From the above, we can use the $\mathcal{STSR}$ to quantify the process of the QST because the perfect QST implies the unchanging $\mathcal{STSR}$ while the imperfect QST decreases the value of the $\mathcal{STSR}$.

\section{Experimental realization}\label{sec:CI}

\begin{figure}[!htpb]
	\centering
	\includegraphics[width=\linewidth]{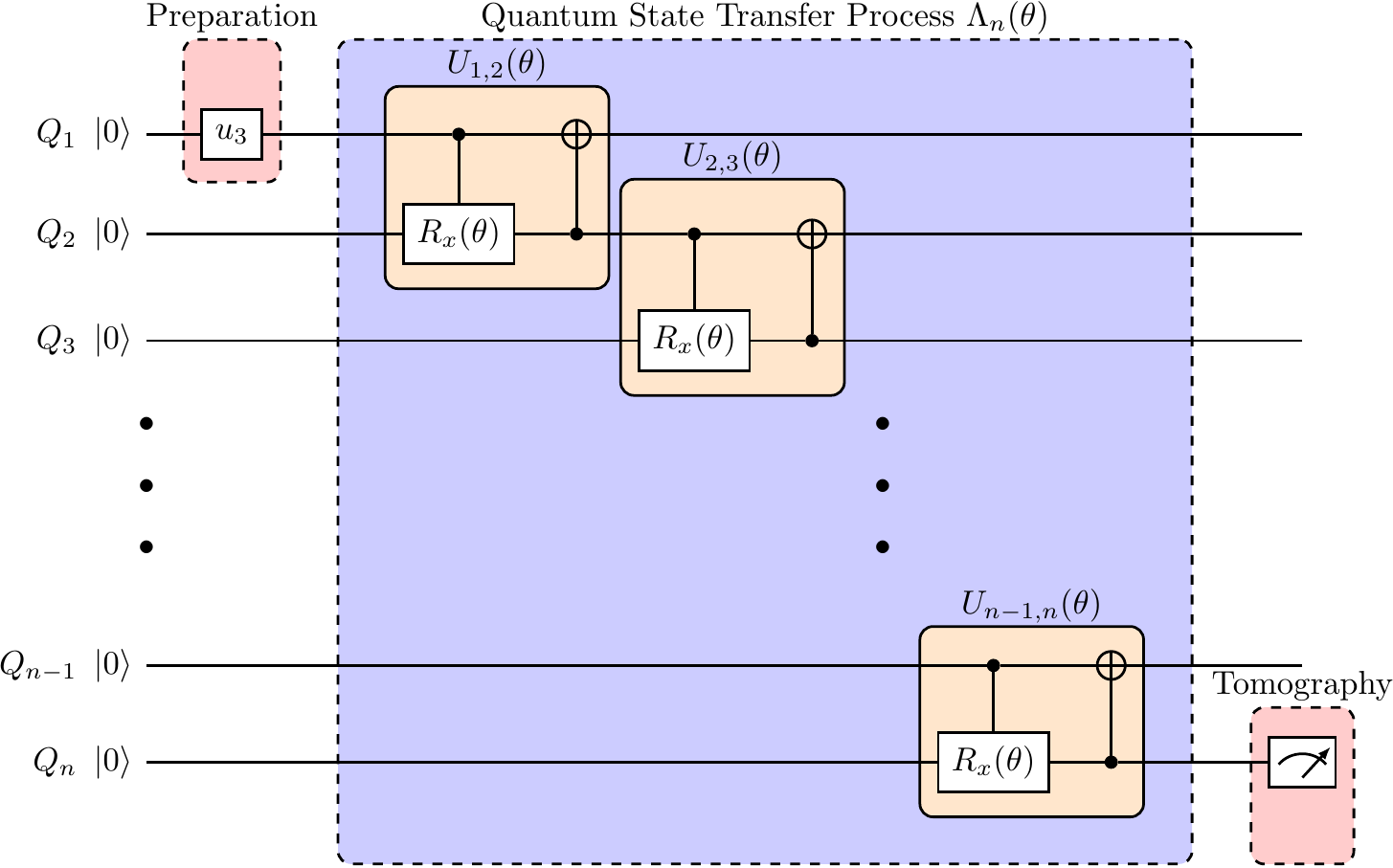}
	\caption{
        Circuit implementation of QST. Here, we prepare the initial states by implementing the $u_3$ operation in qubit $Q_1$. The states can be transferred to the $Q_2$ by applying the $U_{1,2}(\theta)$ operation decomposed with a CRX followed by a CNOT operation. We then iterate the $U_{l,l+1}(\theta)$ operation on the $n$-qubit chain. Finally, the states will be transferred to $Q_n$, which can then be obtained by the procedure of quantum state tomography.
    }
	\label{fig:circuit}
\end{figure}

In this section, we provide a scalable circuit, which can be used to implement the $n$-qubit QST as shown in \fig{\ref{fig:circuit}}. Alice prepares the states in $Q_1$, and Bob receives the transferred states in $Q_n$. To calculate the spatiotemporal steering robustness ($\mathcal{STSR}$), we introduce the preparation method of the assemblage, the quantum state transfer process, and both of their circuit implementations. Moreover, we discuss the ideal theoretical results and model the noise effect by introducing extra qubit decoherence described by the Lindblad master equation.

\subsection{State preparation}\label{sec:state preparation}

%In the STS scenario, Alice and Bob send the post-measurement state and the quantum
%Alice first performs a set of measurements, and send the bi-separable state with Bob %into a global quantum channel. 
Because the IBM quantum experience does not allow one to access the post-measured states after Alice's measurements, we prepare six eigenstates of Pauli matrices being Alice's post-measurement states with indexes $x \in \{1, 2, 3\}$ and $a \in \{1, 2\}$. Note that one can use the ancilla qubit, the CNOT operation, and the measurement operation on the ancilla qubit to replace the measurement operation on the system qubit~\cite{Ku2020-2}. Nevertheless, we consider the state preparation to avoid further errors from the CNOT operation. Note that the gate fidelity and the execution time of the CNOT operation are both at least $10$ times larger than the single qubit operation. Thus, to decrease the errors, the number of CNOT operations should be as less as possible.
The initial state of the qubits on IBM quantum experience is always in $\ket{0}$. We can prepare $\tilde{\varrho}_{a|x}$ by applying the corresponding $u_3(\delta, \phi, \xi)$ operation at $Q_1$, mathematically as follows:
\begin{equation}\label{eq:state_prepare}
\tilde{\varrho}_{a|x} = u_3(\delta, \phi, \xi) \proj{0} u_3^\dagger(\delta, \phi, \xi),
\end{equation}
with the matrix representation of the $u_3(\delta, \phi, \xi)$ operation being
\begin{equation}
u_3(\delta, \phi, \xi) =
    \begin{pmatrix}
        \cos{\frac{\delta}{2}}           & -e^{i \xi}\sin{\frac{\delta}{2}}\\
        e^{i \phi}\sin{\frac{\delta}{2}} &  e^{i(\xi+\phi)}\cos{\frac{\delta}{2}}
    \end{pmatrix}.
\end{equation}
Because we prepare the above states uniformly, $P_\A(a|x) = 0.5~~\forall~~a,x$, and the corresponding assemblage can be obtained by performing Pauli measurements on the maximally mixed state $\id/2$. The above assemblage satisfies the NSIT condition in \eq{\ref{eq:no-signaling}} and can maximize the spatiotemporal steering robustness~\cite{Cavalcanti2016,ShinLiang2020, Ku2018-2}.
 
%cannot apply quantum operations on any given qubits after performing a measurement on it. One of the solution is to perform a CNOT operation on the measured qubit(control) and an ancilla(target) qubit. The other solution we choosed is to prepare the state directly.  In order to quantify nonclassicality of QST by calculating spatiotemporal steering robustness ($\mathcal{STSR}$), we prepare a set of states $\assemb{\tilde{\varrho}_{a|x}}$, with \eq{\ref{eq:state_prepare}}, to be the six eigenstates of Pauli matrices, that is, $x \in \{X, Y, Z\}$ and $a \in \{0, 1\}$. Also, to make the assemblage $\assemb{\varrho_{a|x}}$ satisfy NSIT condition in \eq{\ref{eq:no-signaling}}, we assume
%\begin{equation}
%P_\A(a|x) = 0.5 ~~\forall~~ a,~x.
%\end{equation}
%\comt{
%\begin{center}
%\begin{tabular}{ c c c }
%    $\varrho_{-|X} =$ &$u_3(\frac{\pi}{2}, \pi, \pi) \proj{0} u_3^\dagger(\frac{\pi}{2}, \pi, \pi)$ &,\\ 
%    $\varrho_{+|X} =$ &$u_3(\frac{\pi}{2}, 0, \pi) \proj{0} u_3^\dagger(\frac{\pi}{2}, 0, \pi)$ &,\\
%    $\varrho_{-|Y} =$ &$u_3(\frac{\pi}{2}, \frac{-\pi}{2}, \pi) \proj{0} u_3^\dagger(\frac{\pi}{2}, \frac{-\pi}{2}, \pi)$ &,\\ 
%    $\varrho_{+|Y} =$ &$u_3(\frac{\pi}{2}, \frac{\pi}{2}, \pi) \proj{0} u_3^\dagger(\frac{\pi}{2}, \frac{\pi}{2}, \pi)$ &,\\
%    $\varrho_{-|Z} =$ &$\proj{0}$ &,\\ 
%    $\varrho_{+|Z} =$ &$u_3(\pi, 0, \pi) \proj{0} u_3^\dagger(\pi, 0, \pi)$ &.
%\end{tabular}
%\end{center}
%}

\subsection{Quantum state transfer process}\label{sec:quantum state transfer process}

\begin{figure}[!htpb]
  \includegraphics[width=\linewidth]{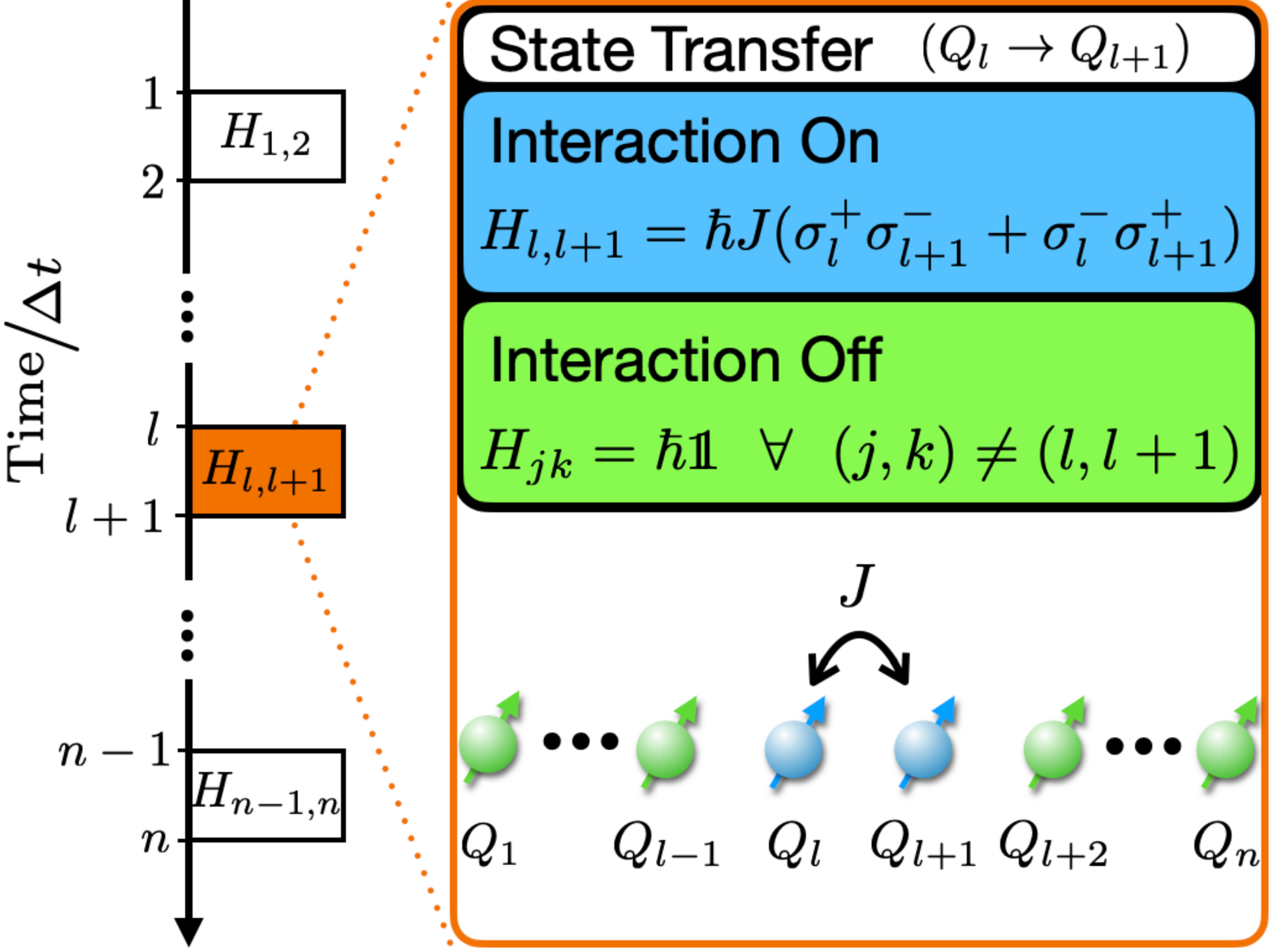}
\caption{
      Schematic illustration of the perfect $n$-qubit QST process $\Lambda_t$, which consists of several iterations of quantum operations. For the $l$th iteration, the state is transferred from $Q_l$ to $Q_{l+1}$ by turning on the qubit-qubit interaction [depicted in Eq.~\eqref{eq:hamiltonian} and Eq.~\eqref{eq:time evolution}] for a period of time $\Delta t=\pi/(2J)$, where $J$ is $1/2$. Therefore, to transfer Alice's prepared states from $Q_1$ to $Q_n$, the iteration has to be performed $n-1$ times.
      % Here, $S$ is just an unitary operation and the state can be recovery by applying the inverse unitary operation. 
  }
\label{fig:interaction}
\end{figure}

We consider a QST process described by an $n$-qubit chain, as shown in \fig{\ref{fig:interaction}}, with each qubit labeled as $Q_l$, where $l=1,2, \dots ,n$. In this process, Alice prepares the states in $Q_1$, and after the QST process, Bob will receive the transferred states in $Q_n$. We consider a QST procedure, which involves several iterations of quantum operations. For each iteration, we turn on the qubit-qubit interaction between $Q_l$ and $Q_{l+1}$, and then, turn it off when the QST from $Q_l$ to $Q_{l+1}$ is accomplished. Here, the ``closed" interaction can be represented by the identity operator in the interaction Hamiltonian. The interaction Hamiltonian between $Q_l$ and $Q_{l+1}$~\cite{Li2018,Shin-Liang2017} is written as  
\begin{equation}\label{eq:hamiltonian}
	H_{{l,l+1}} = \hbar J(\sigma_{l}^+\sigma_{l+1}^- + \sigma_{l}^-\sigma_{l+1}^+),
\end{equation}
where $J$ is the coupling strength between $Q_l$ and $Q_{l+1}$. $\sigma_{l}^+ (\sigma_{l}^-)$ is the raising (lowering) operator acting on $Q_l$. Without loss of generality, $J$ can be set to $1/2$. The corresponding time evolution unitary operator can then be written as 
\begin{align}\label{eq:time evolution}
    \mathcal{V}_{l,l+1}(t)  &=\exp(-iH_{l,l+1}t/\hbar) \nonumber\\
    %e^{-iH_{\tiny{\text{A,B}}}t/\hbar} =
	&=\begin{pmatrix}
		1 &            0 &            0 & 0 \\
		0 &   \cos{\frac{t}{2}} & -i\sin{\frac{t}{2}} & 0 \\
		0 & -i\sin{\frac{t}{2}} &   \cos{\frac{t}{2}} & 0 \\
		0 &            0 &            0 & 1 
    \end{pmatrix}_{l,l+1},
\end{align}
where the matrix representation of the unitary operator is expanded in the computational basis for $Q_l$ and $Q_{l+1}$. %namely $\{|m_l,n_{l+1}\rangle\langle p_l,q_{l+1}|\}_{m,n,p,q\in \{0,1\}}$.
Therefore, when the two qubit state is initialized in $\rho\otimes |0\rangle\langle0|$, the reduced state $\rho^\prime$ for $Q_{l+1}$ after the evolved time $\Delta t=\pi$ reads as follows:
\begin{equation}\label{eq:QST two qubit}
    \rho^\prime = \tr_{l}[~\mathcal{V}_{l,l+1}(\Delta t)~ (\rho \ts \proj{0})~ \mathcal{V}_{l,l+1}^\dagger(\Delta t)~] = S^\dagger \rho S,
\end{equation}
where $S=\text{diag}(1,i)$ is a unitary operator. Obviously, the state $\rho$ is perfectly transferred from $Q_l$ to $Q_{l+1}$ because the fidelity between the prepared and received states under the decoding unitary operation $S$ is unity. We note that the effective dynamics of the $\mathcal{V}_{l,l+1}(\Delta t)$ is identical to the $i$SWAP$^\dagger$ operation. If one of the subsystems is $\ket{0}$($\ket{1}$), the $i$SWAP$^\dagger$ operation can be viewed as a SWAP operation together with a $S^\dagger$($S$) operation. Also note that while considering the $XYZ$ interaction Hamiltonian in Ref.~\cite{Kandel2019}, the evolution operator is proportional to the SWAP operation.

Accordingly, to transfer Alice's prepared states from $Q_1$ to $Q_n$, $n-1$ times of the aforementioned two-qubit operations are required. The total QST process can be described using the following unitary operation
\begin{equation}\label{eq:total time evolution}
  \tilde{\mathcal{V}}_{1, n} = \prod_{l=n-1}^{1} \mathcal{V}_{l, l+1} ~,
\end{equation}
Finally, the unitary $S^{n-1}$ is applied on $Q_n$, and Bob obtains the states that are the same as Alice's prepared states. However, based on \eq{\ref{eq:PST_SR}}, the decoding unitary operation $S^{n-1}$ is unnecessary when considering the STS scenario. Usually, for digital quantum processors, e.g. the IBM quantum experience and QuTech quantum inspire considered in this work, the $S^{n-1}$ operation comprises a sequence $S$-gate.
%Therefore, getting rid of these operations dramatically decreases the circuit depth of the process when $n$ is large; i.e. the accumulated errors from the sequence of operations are reduced.

\begin{figure}[!htbp]
    \subfloat[$\mathcal{V}_{l, l+1}(\theta)$ \label{fig:decomposition_theory}]{\includegraphics[width=.5\linewidth]{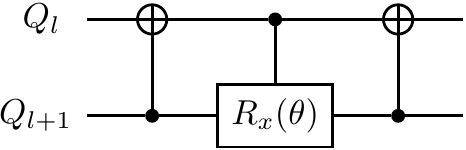}}~~~
    \subfloat[$U_{l, l+1}(\theta)$ \label{fig:decomposition_exp}]{\includegraphics[width=.41\linewidth]{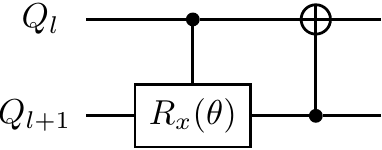}}
    \caption{Circuit decomposition of $\mathcal{V}_{l,l+1}(\theta)$ and $U_{l,l+1}(\theta)$. $R_x(\theta)$ is the rotation $X$ operation with rotating angle $\theta$.}
    \label{fig:unitary_decomposition}
\end{figure}

The circuit implementation of the evolution operator $\mathcal{V}_{l,l+1}(\theta)$ in \eq{\ref{eq:time evolution}} is shown in \fig{\ref{fig:decomposition_theory}}. Here, we replace $t$ with $\theta$. To implement the controlled rotation $X$ (CRX) in IBM quantum experience, one has to decompose it with two CNOT operations and three $u_3$ operations. Thus, there are a total of four CNOT operations in the evolution operator $\mathcal{V}_{l,l+1}(\theta)$. As mentioned above, we would like to decrease the number of the CNOT operations to decrease the inevitable errors. Thus, we consider an alternative unitary operator $U_{l,l+1}(\theta)$, which reduces one CNOT operation, as 
\begin{equation}\label{eq:u}
U_{l,l+1}(\theta) = 
\begin{pmatrix}
    1 & 0 & 0 & 0 \\
    0 & 0 & -i\sin{\frac{\theta}{2}} & \cos{\frac{\theta}{2}} \\
    0 & 0 & \cos{\frac{\theta}{2}}   & -i\sin{\frac{\theta}{2}} \\
    0 & 1 & 0 & 0
\end{pmatrix}_{l,l+1},
\end{equation}where the circuit implementation of $U_{l,l+1}$ is shown in \fig{\ref{fig:decomposition_exp}}.
If the initial state in $Q_{l+1}$ is always $\proj{0}$, we can replace $\mathcal{V}_{l,l+1}$ with $U_{l,l+1}$; thus, Eq.~\eqref{eq:QST two qubit} still holds, where
\begin{alignat}{1}\label{eq:rho_B_exp}
    \rho^\prime &= \tr_{l}[\mathcal{V}_{l,l+1}(\theta) (\rho \ts \proj{0}) \mathcal{V}_{l,l+1}^\dagger(\theta)] \notag \\
    &= \tr_{l}[U_{l,l+1}(\theta) (\rho \ts \proj{0}) U_{l,l+1}^\dagger(\theta)].
\end{alignat}
For this implementation, the QST process is perfect only when $\theta=\pi$, that is, $\rho$  and $\rho^\prime$ are related by unitary operation $S$. We refer to the cases where $\theta\neq\pi$ as imperfect QST processes because the transferred states cannot be transformed to the prepared states through an decoding unitary transformation.

\subsection{Ideal theoretical results}

\begin{figure}[!htbp]
  \includegraphics[width=\linewidth]{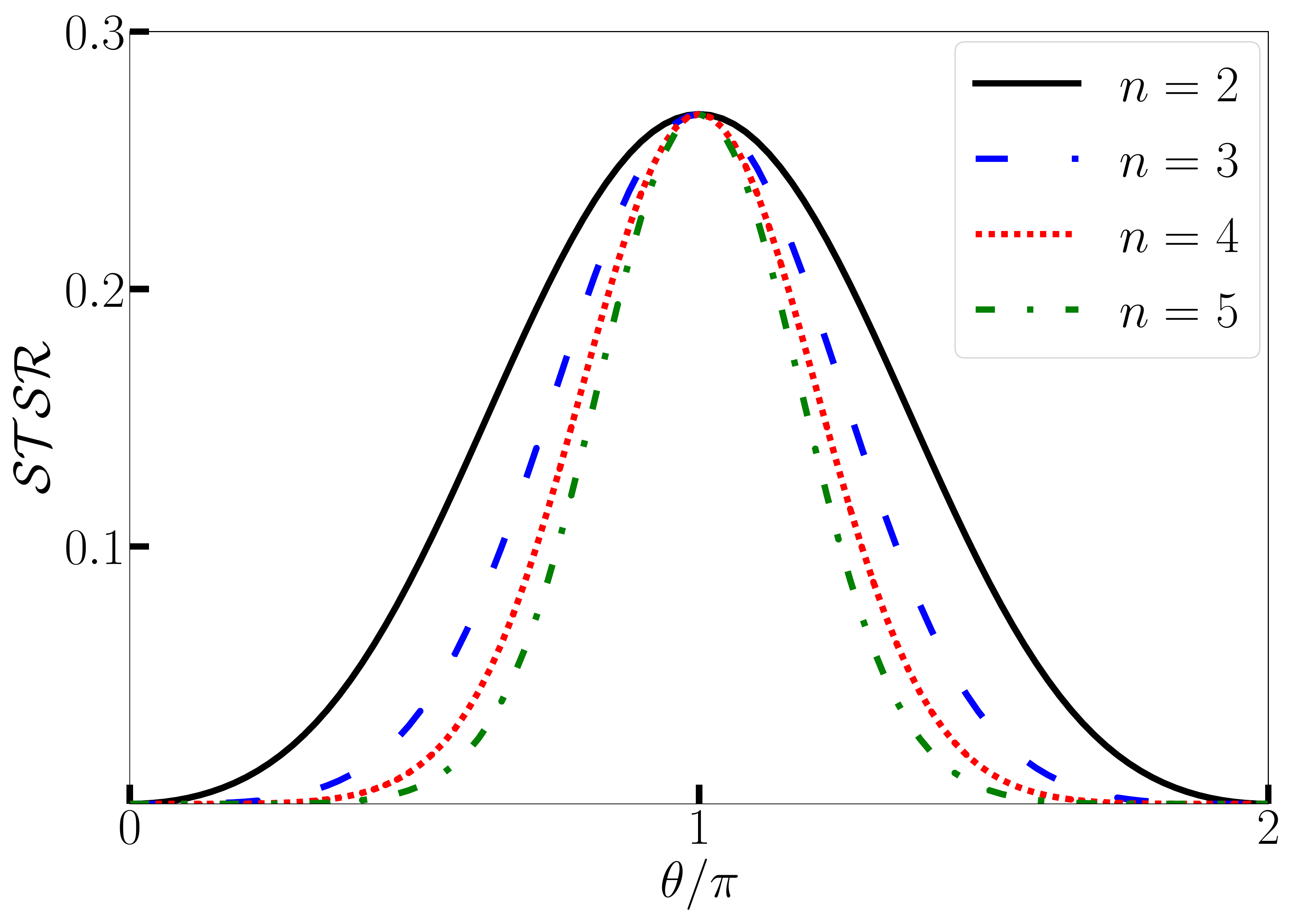}
  \caption{Value of $\mathcal{STSR}$ with respect to different angle $\theta$ for different qubit number $n$. Here, the initial assemblage for $Q_1$ is   $\assemb{\varrho_{a|x} = P_{\A}(a|x)\tilde{\varrho}_{a|x}}_{a,x}$ where $P_\A(a|x)=1/2$ for all $a,x$ and $\tilde{\varrho}_{a|x}$ are the eigenstates of Pauli operators.}
  \label{fig:theory}
\end{figure}

Figure~\ref{fig:theory} shows theoretical predictions of $\mathcal{STSR}$ with respect to the parameter $\theta$ for different qubit numbers $n\in\{2,3,4,5\}$. We can observe that for fixed $n$, the value of $\mathcal{STSR}$ for the perfect QST case ($\theta=\pi$) is always larger than those for the imperfect QST cases ($\theta\neq\pi$). This is because for a fixed $n$, the assemblages for the $\theta=\pi$ case and those for the $\theta\neq\pi$ cases are, in general, related by a unitary transformation and a completely positive and trace-preserving map (CPTP), respectively. It has been proved that $\mathcal{STSR}$ monotonically decreases whenever the underlying assemblage is sent into a CPTP map~\cite{Rodrigo2015}.

Moreover, for fixed $\theta$, the value of the $\mathcal{STSR}$ monotonically decreases with increasing qubit number $n$. As shown in Figs.~\ref{fig:circuit} and \ref{fig:interaction}, increasing $n$ means increasing the number of the iterations required in the QST process. As described in Eq.~\eqref{eq:rho_B_exp}, for each iteration, the input state $\rho$ and the output state $\rho^\prime$ can also be generally related by a CPTP map. Therefore, when increasing the number $n$, the prepared assemblage will be iteratively sent into the CPTP maps, which results in a decrease of the $\mathcal{STSR}$~\cite{Rodrigo2015}. 

%Especially, when $\theta=0$, the evolution operator $U(0)$ is the same as the CNOT operation with the transferred and received states being the target and control qubits, respectively. Since the received state is a control qubit, the state cannot be transferred.

%\fig{\ref{fig:theory}} shows the theoretical predictions of the $\mathcal{STSR}$ for different $n$. From this figure, we observe the invariance of $\mathcal{STSR}$ for different $n$ when $\theta=\pi$, which means that perfect QST only succeed when $\theta=\pi$. For $\theta=0$, since $U(0)$ is identical to the CNOT operation but with $Q_\B$ ($Q_\A$) being the control (target) qubit, respectively. 

\subsection{Noise simulation}\label{sec:noise_simulation}

Because the quantum devices nowadays suffer from noise due to the interactions with environments~\cite{Johansson2012,Johansson2013}, we model the noise effect by introducing extra qubit decoherence (dephasing and relaxation) described by the following Lindblad master equation (similar discussions can be found in the Ref.~\cite{Ku2020-2}):
\begin{equation}
\begin{aligned}
\dot{\rho}(t) &= \sum_l^n \hbar\frac{\gamma^l_{T_1}}{2}\left[ 2\sigma_{-}^{l}\rho(t)\sigma_{+}^{l} - \sigma_{+}^{l}\sigma_{-}^{l}\rho(t) - \rho(t)\sigma_{+}^{l}\sigma_{-}^{l} \right]\\
&+ \sum_l^n \hbar\frac{\gamma^l_{T_2}}{2}\left[ 2\sigma_{z}^{l}\rho(t)\sigma_{z}^{l} - {\sigma_{z}^{l}}^2\rho(t) - \rho(t){\sigma_{z}^{l}}^2 \right],
\end{aligned}
\label{eq:master}
\end{equation}
where $\rho$ denotes the density operator for the total n-qubit system, and the coefficients $\gamma^l_{T_1}=1/T_1^l$ and $\gamma_{T_2}^l=(1/T_2^l - 1/2T_2^l)/2$ are the qubit relaxation and decoherence rates for $Q_l$, respectively. Here, $T_1^l$ and $T_2^l$ represent the relaxation and dephasing time for $Q_l$, respectively. The relaxation and the coherence time for each qubit and the operation-execution time are all public in IBM quantum experience~\cite{IBMQ}. We use these public information together with the master equation to model the decoherence effect for real devices, such that the final reduced state $\rho_n$ of $Q_n$ can be obtained when tracing out other qubits.

We further consider the measurement (readout) errors, which is not described in the aforementioned master equation. To insert such errors, we briefly recall how to obtain measurement errors in IBM quantum experience. In the measurement-error calibration, we always measure the system in computational basis while initializing the qubit with two basis states, $\ket{0}$ and $\ket{1}$. For the ideal situation, the measurement outcome is $0$ ($1$) with certainty when the qubit is initialized in $|0\rangle$ ($|1\rangle$). Therefore, measurement errors $\Gamma$ can be determined by the average probability of preparation in $\ket{0} (\ket{1})$ with 
the opposite outcome $1$ ($0$). We model such errors by sending the quantum state into the bit-flip channel before measurement; i.e., 
\begin{equation}\label{eq:readout}
\rho_n\rightarrow (1-\Gamma) \rho_n+\Gamma X\rho_nX^\dagger.
\end{equation}
The above channel changes the population of the quantum state $\rho_n$ with probability $\Gamma$. Notably, once the state is $\ket{0}$ or $\ket{1}$, the population obtained by the above is the same as the one in the measurement-error calibration.
Finally, we emphasize that since both \eq{\ref{eq:master}} and \eq{\ref{eq:readout}} can be described by CPTP maps, $\mathcal{STSR}$ can only decrease~\cite{Rodrigo2015} after introducing our noisy model.

\begin{figure*}[!htbp]
    \includegraphics[width=0.75\linewidth]{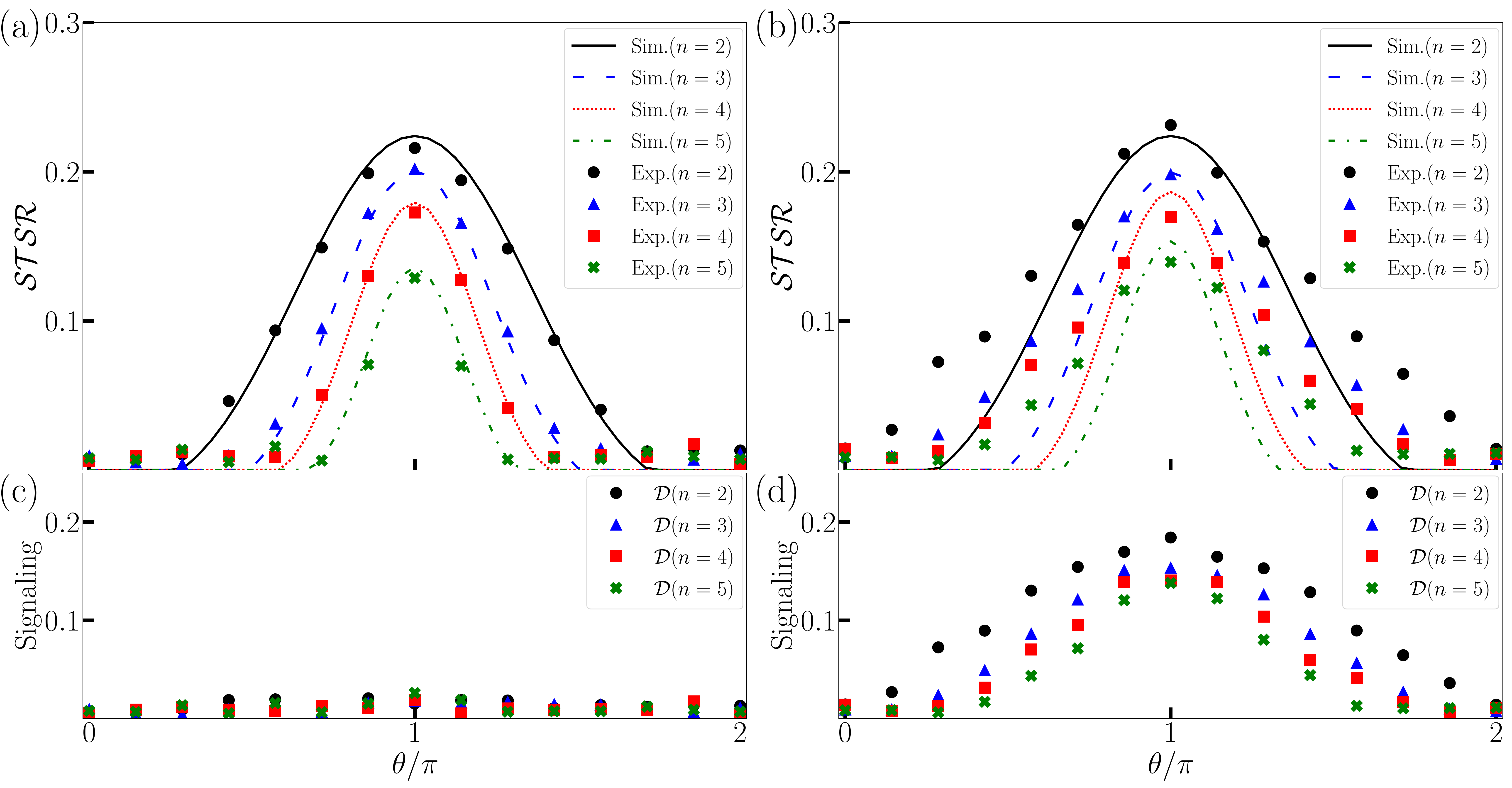}
    \caption{
        The experimental values of the $\mathcal{STSR}$ under the QST. The evolution operators transfer the state from $Q_1$ to $Q_n$. The experimental results of (a) and (c) are implemented in Mar 2020 after maintenance, whereas those for (b) and (d) were complete in Jan 2020 before the maintenance. In (a) and (b), we show the spatiotemporal steerability as a quantification of the QST process. When $\theta=\pi$, the ideal evolution circuit, corresponding to the perfect QST process, provides the maximal spatiotemporal steerability. Although the experimental $\mathcal{STSR}$ cannot reach the theoretical prediction, the trend of the experimental results functions similarly with the ideal case. Moreover, one can observe that the $\mathcal{STSR}$ well fitted the noisy model for (a) and (b). One can see that the signaling effect in (d) actually dominates $\mathcal{STSR}$ in (b) by \eq{\ref{eq:signaling}}.
    }
    \label{fig:exp}
\end{figure*}

\section{Experimental Results}\label{sec:result}

We prepare the eigenstates of Pauli matrices in $Q_1$ using the corresponding single-quantum operation $u_3$, which rotates the $|0\rangle$ to the prepared states (see \sect{\ref{sec:state preparation}}). The global evolution $\Lambda_n(\theta)$ is then applied as shown in \fig{\ref{fig:circuit}} with different qubit numbers $n \in \{2, 3, 4, 5\}$. After sending the system into the channel $\Lambda_n(\theta)$, we can reconstruct the reduced density matrices on $Q_n$ by standard state tomography. Here, the measurement results are obtained through $8000$ shots for each procedure in the state tomography.

In \fig{\ref{fig:exp}}, we present experimental data with $\theta\in\{\frac{m\pi}{7}|m=0, 1, ... , 14\}$ obtained from different dates on March 2020 and January 2020 for the same device named IBMQ boeblingen. The experiment shown in \fig{\ref{fig:exp}} (a) and (c) was completed right after maintenance, whereas the results in \fig{\ref{fig:exp}} (b) and (d) were obtained before the maintenance. We also provide the noise simulation mentioned in \sect{\ref{sec:noise_simulation}} and the violation of the NSIT described in Eq.~\eqref{eq:signaling}. One can find that the value of the $\mathcal{STSR}$ at $\theta=\pi$, where the perfect QST occurs for the ideal case, decreases as the qubit number $n$ increases. The reduction agrees with the noise simulations, suggesting that the QST nonclassicality is suppressed because of accumulation of noise. Additionally, It seems that the overall $\mathcal{STSR}$ or the QST nonclassicality for the result before the maintenance are larger than that after maintenance. However, by observing \fig{\ref{fig:exp}}(c) and \fig{\ref{fig:exp}}(d), we can clearly find that the $\mathcal{STSR}$ before the maintenance, as shown in \fig{\ref{fig:exp}}(b), is actually dominated by the signaling effect, which cannot be regarded as a genuine quantumness. Therefore, benchmarking nonclassicality of a quantum device requires both $\mathcal{STSR}$ and the condition of NSIT.

Furthermore, there exists intrinsic non-Markovianity in the quantum processors~\cite{Morris2019,Harper2019}. The non-Markovianity is a possible source of the violation of NSIT shown in the presented experimental results because the existence of the non-Markovian effect implies that the operations shown in \fig{\ref{fig:circuit}} could not be divisible. In other words, the global evolution $\Lambda_n(\theta)$ could depend on the state preparation operation $u_3$ and could result in the violation of Eq.~\eqref{eq:no-signaling}.

Finally, the experimental results of the perfect QST from the other quantum devices based on spin qubits (QuTech spin-2) in silicon~\cite{Vandersypen2017,Watson2018} and the superconducting transmon qubits (QuTech starmon-5) are also presented in \tab{\ref{tab:result compare}}. Because QuTech devices do not support the generalized $u_3$ and CNOT operation (one has to decompose the arbitrary quantum operations by a serious single quantum operations and the CZ operations), we only consider the perfect QST case which can be decomposed by $H, S, Z$, and CNOT operations. The qubit relation time ($T_1$) and the coherence time ($T_2$) given by IBM quantum experience~\cite{IBMQ} is about six to eight times longer than those given by QuTech quantum inspire~\cite{Qutech}. Generally speaking, IBM quantum experience outperforms QuTech quantum inspire when both the $\mathcal{STSR}$ and singalling effect are considered. This could be because of the unwanted operation decompositions on the CRX and the CNOT operation such that the noise effect and circuit depth increase. The signaling effect in QuTech spin-2 dominates the result, just like IBMQ's results before maintenance. We also present $\mathcal{STSR}$ under the perfect QST process on other IBM quantum devices (in \apx{\ref{App_diff_IBM_chips}}).
\begin{table}[!htbp]
  \caption{
      The values of the $\mathcal{STSR}$ under the perfect QST process with different quantum devices. Here, we consider the IBMQ boeblingen, QuTech starmon-5, and spin-2 systems. The transference route initially begins with the preparation in $Q_0$, passing through the intermediary qubits, and performs state tomography on the final qubit. The experimental results of the signaling effect are also presented.
  }
  \label{tab:result compare}
  \begin{tabular}{ clccc }
      \hline
      \hline
      Devices & Transference routes & $~n~$ & $~\mathcal{STSR}~$ & Signaling \\
      \hline
      \multirow[t]{4}{*}{\vtop{\hbox{\strut IBMQ}\hbox{\strut boeblingen}\hbox{\strut (Mar, 2020)}}} & \multirow[t]{4}{*}{$0 \to 1 \to 2 \to 3 \to 4$} & 2 & 0.216 & 0.015 \\
      & & 3 & 0.202 & 0.018 \\
      & & 4 & 0.173 & 0.019 \\
      & & 5 & 0.129 & 0.026 \\
      \hline
      \multirow[t]{3}{*}{\vtop{\hbox{\strut QuTech}\hbox{\strut starmon-5}\hbox{\strut (May, 2020)}}} & \multirow[t]{3}{*}{$0 \to 2 \to 4$} & 2 & 0.170 & 0.051 \\
      & & 3 & 0.054 & 0.035 \\
      &&&&\\
      \hline
      \multirow[t]{3}{*}{\vtop{\hbox{\strut QuTech}\hbox{\strut spin-2}\hbox{\strut (May, 2020)}}} & $0 \to 1$ & 2 & 0.103 & 0.100\\
      &&&&\\
      &&&&\\
      \hline
      \hline
  \end{tabular}
\end{table}

\section{Discussion}\label{sec:conclusion}

In this work, we first defined the classical strategy of state transfer as the received state can be constructed by an ensemble of ontic states together with a stochastic map. Such a strategy can be described by the local-hidden-state model which is widely used in the context of the steering scenarios. We then proposed a method based on STS to quantify the nonclassicality of the QST process. We have shown that the spatiotemporal steerability is invariant under the process of the perfect QST, whereas the reduction during the process of the QST is imperfect. Moreover, we have provided a quantity to estimate signaling effect and proved that such a quantity is a lower bound of $\mathcal{STSR}$.%We note that in general the classical process of the state transfer can be described by the measurement-and-prepare channel and it cannot reveal the spatiotemporal steerability. 

Not only did we realize a proof-of-principle experiment but also performed a benchmark experiment of the QST process on IBM quantum experience and QuTech quantum inspire. 
Our experimental results show that the degrees of QST nonclassicality decrease as the circuit depth increases. In addition, the decrease agrees with the noise model, which describes the accumulation of noise (qubit relaxation, gate error, and readout error) during the QST process.
The experimental results from the IBMQ boeblingen before the maintenance shows that the spatiotemporal steerability is actually dominated by the signaling effect. Such signaling effect could be caused by the intrinsic non-Markovianity for the quantum devices.

In Ref.~\cite{Bayat2010}, it has been shown that the average fidelity of the QST process is identical to estimate the degree of entanglement distribution. More specifically, by keeping one part of the maximally entanglement pair in the sender and sending the other part to receiver through QST process, we can compute the singlet fraction of the output state ~\cite{Horodecki1999}. The above approach has been used to quantify the QST process~\cite{Horodecki1999}. In general, this output state is the Choi state, which is the one-to-one mapping between the quantum state and quantum channel. We recall that the degree of the entanglement can be estimated by spatial steerability~\cite{Uola2020, Ku2020, ShinLiang2020, Piani2015}. Furthermore, violating the steering inequality is related to the singlet fraction~\cite{Hsieh2016}. Due to the hierarchy relation in Ref.~\cite{Ku2018-2}, STS can access partial information of the Choi state. Therefore, it is naturally to ask whether spatiotemporal steerability can estimate the amount of entanglement distribution.

Throughout this work, although we only consider the single-qubit QST, we briefly discuss the $d$-level~\cite{Bayat2014, Liu2017} and multi-partite QST~\cite{Lorenzo2013}. Since, in \eq{\ref{eq:QST_rho_B}}, the dimension of the prepared and received states can be arbitrary, our approach can be easily extended to $d$-level QST. It would be more interesting to consider multi-partite QST. In such a scenario, depending on the structure of the assemblage, one could introduce more constraints on the ontic states in LHS model~\cite{Cavalcanti2015}. For instance, consider the case where the assemblage contains two sets of two-qubit entangled states, there are at least two different ways to define the ontic states in LHS model: One could either allow the ontic states to be arbitrary or separable two-qubit states. Therefore, the nonclassicality of multi-partite QST could be very versatile (e.g., the ontic states could be $m$-separable in the notion of genuine multi-partite entanglement~\cite{Cavalcanti2015,Zhou2019,Lu2018, G_hne2009}).

This work also raises some open questions. Can we characterize the non-Markovian effect? Can we implement the QST process with less CNOT operations? In our work, we have to use three CNOT operations to implement QST, while the number of the CNOT operations is the same as the operation decomposition of the SWAP operation.

\section*{Acknowledgements}
We acknowledge the NTU-IBM Q Hub (Grant: MOST 107-2627-E-002-001-MY3) and the IBM quantum experience for providing us a platform to implement the experiment. The views expressed are those of the authors and do not reflect the official policy or position of IBM or the IBM Quantum Experience team.
The authors acknowledge fruitful discussions with Al\'an Aspuru-Guzik, Neill Lambert, Gelo Noel Tabia, Shin-Liang Chen, and Po-Chen Kuo. In particular, we thank Gelo Noel Tabia for his insightful discussion on the proof of estimating signaling effect.
The authors acknowledge the support of from the National Center for Theoretical Sciences and Ministry of Science and Technology, Taiwan (Grants Nos. MOST 107-2628-M-006-002-MY3, 109-2627-M-006-004), the National Center for Theoretical Sciences and Ministry of Science and Technology, Taiwan (Grant No. MOST 108-2811-M-006-536) for H.-Y.K., and Army Research Office (Grant No. W911NF-19-1-0081) for Y.-N.C.

% Start appendix
\appendix
\section{Semidefinite programming for spatiotemporal steering robustness}\label{sec:SDP}

Here, we briefly describe the semidefinite program (SDP) of the spatiotemporal steering robustness $\mathcal{STSR}$ in \eq{\ref{eq:robustness}} which is first introduced in~\cite{Shin-Liang2017}. We also note that the SDP of the $\mathcal{STSR}$ is identical to that in the spatial and temporal steering scenarios~\cite{Piani2015,Ku2016,Cavalcanti2016}. 

Let us consider $m$-measurement settings $x \in \assemb{1, 2, ..., m}$ where each has $q$ outcomes $a \in \assemb{1, 2, ..., q}$. Since inputs and outcomes are finite, the number of the variable $\lambda$ in \eq{\ref{eq:LHS}} is $q^m$. Each $\lambda$ can be considered as a string of ordered outcomes according to the measurements: $(a_{x=1}, a_{x=2}, ..., a_{x=m})$. We can define the deterministic strategy function $D_\lambda(a|x)=\delta_{a, \lambda(x)}$, where $\delta$ is the Kronecker delta function and $\lambda(x)$ denotes the value of the string at position $x$~\cite{Uola2020, Cavalcanti2016}. Therefore, given an assemblage $\assemb{\varrho_{a|x}}$, the primal SDP of
$\mathcal{STSR}$ can be formulated as follows (see the derivation in Refs.~\cite{Cavalcanti2016, Uola2020}):
\begin{alignat}{2}\label{eq:SDP}
    {\displaystyle \min_{\assemb{\sigma_\lambda}}} &~&& \tr\sum_\lambda \sigma_\lambda - 1,\notag\\
    \text{s.t.} &~&&\sum_\lambda D_\lambda(a|x)\sigma_\lambda - \varrho_{a|x} \geq 0 ~~ \forall ~ a, x~, \notag\\
    &~&&\sigma_\lambda \geq 0  ~~ \forall ~ \lambda ~ .
\end{alignat}
The dual formulation of \eq{\ref{eq:SDP}} is given by~\cite{Cavalcanti2016, Uola2020}
\begin{alignat}{2}\label{eq:dual_SDP}
    {\displaystyle \max_{\assemb{F_{a|x}}}} &~&& \tr\sum_{a,x} F_{a|x}\varrho_{a|x} - 1, \notag\\
    \text{s.t.} &~&& \id - \sum_{a,x} D_\lambda(a|x) F_{a|x} \geq 0 ~~ \forall ~ \lambda ~, \notag\\
    &~&& F_{a|x} \geq 0  ~~ \forall ~ a,x.
\end{alignat}
Here, $F_{a|x}$ is the steering witness that distinguishes the steerable assemblage from the unsteerable ones. We note that the strong duality of $\mathcal{STSR}$ has been shown in Refs.~\cite{Clemente2015, Cavalcanti2016}, meaning that the results of the primal and dual formulations are equivalent.

\section{Proof of \eq{\ref{eq:PST_SR}} in the main text}\label{sec:proof_invariant}

In this section, we show that given an assemblage $\assemb{\varrho_{a|x}}$, the $\mathcal{STSR}$ is invariant under unitary transformation using the strong duality mentioned in \apx{\ref{sec:SDP}}. More specifically, we show $\mathcal{STSR}(\assemb{\varrho_{a|x}'}) = \mathcal{STSR}(\assemb{\varrho_{a|x}})$, where $\varrho_{a|x}' = U\varrho_{a|x}U^\dagger$ with $U$ being an arbitrary unitary operator. 

Because the dual formulation of SDP in \eq{\ref{eq:dual_SDP}} of $\mathcal{STSR}$ is strongly feasible, given an assemblage $\assemb{\varrho_{a|x}}$, one can always find the optimal spatiotemporal steering witness $\assemb{F_{a|x}^*}$ satisfying both constraints in \eq{\ref{eq:dual_SDP}}:
\[
    \mathcal{STSR}(\assemb{\varrho_{a|x}}) = \tr\sum_{a,x} F_{a|x}^*\varrho_{a|x} -1 \notag.
\]

With the above, we now apply a unitary transformation $U$ on the given assemblage $\assemb{\varrho_{a|x}}$. The dual formulation of $\mathcal{STSR}(\assemb{\varrho_{a|x}'})$ can be expressed as follows:
\begin{alignat}{1}
    \mathcal{STSR}(\assemb{\varrho_{a|x}'})&={\displaystyle \max_{\assemb{F_{a,x}'}}} ~ \tr\sum_{a,x} F_{a|x}'\varrho_{a|x}' - 1 \notag \\
    & \geq \tr\sum_{a,x} (UF_{a|x}^*U^\dagger) (U\varrho_{a|x}U^\dagger) - 1 \notag\\
    & = \tr\sum_{a,x} UF_{a|x}^*\varrho_{a|x}U^\dagger - 1 \notag\\
    & = \tr\sum_{a,x} F_{a|x}^*\varrho_{a|x} - 1 \notag\\
    & = \mathcal{STSR}(\assemb{\varrho_{a|x}}) ~. \notag
\end{alignat}
The inequality holds because $\assemb{UF_{a|x}^*U^\dagger}$ is not the optimal solution of SDP. Nevertheless, it is indeed a valid solution because it satisfies both constraints in \eq{\ref{eq:dual_SDP}}:
\begin{alignat}{2}
    &\id - \sum_{a,x} D_\lambda(a|x) F_{a|x}' &&= \id - \sum_{a,x} D_\lambda(a|x) UF_{a|x}^*U^\dagger \notag\\
    &~&&=U\left(\id - \sum_{a,x} D_\lambda(a|x) F_{a|x}^*\right)U^\dagger \notag\\
    &~&&\geq 0 ~~ \forall ~ \lambda ~, \notag\\
    &F_{a|x}' = UF_{a|x}^*U^\dagger \geq && 0  ~~ \forall ~ a,x. \notag
\end{alignat}
Therefore, we arrive at the bound relation; i.e., 
\begin{equation}\label{eq:stsr_lower}
    \mathcal{STSR}(\assemb{\varrho_{a|x}'}) \geq \mathcal{STSR}(\assemb{\varrho_{a|x}}).
\end{equation}

A similar argument can also be applied to the primal SDP in \eq{\ref{eq:SDP}} of $\mathcal{STSR}$. Given an assemblage, one can always find the optimal $\assemb{\sigma_\lambda^*}$ that satisfies both constraints in \eq{\ref{eq:SDP}}:
\[
    \mathcal{STSR}(\assemb{\varrho_{a|x}}) = \tr\sum_\lambda \sigma_\lambda^* -1.
\]

By applying a unitary transformation $U$ on the given assemblage $\assemb{\varrho_{a|x}}$, the primal SDP of $\mathcal{STSR}(\assemb{\varrho_{a|x}'})$ can then be expressed as follows:
\begin{alignat}{1}
    \mathcal{STSR}(\assemb{\varrho_{a|x}'})&={\displaystyle \min_{\assemb{\sigma_\lambda'}}} ~ \tr\sum_\lambda \sigma_\lambda' - 1 \notag \\
    & \leq \tr\sum_\lambda U\sigma_\lambda^*U^\dagger - 1 \notag\\
    & = \tr\sum_\lambda \sigma_\lambda^* - 1 \notag\\
    & = \mathcal{STSR}(\assemb{\varrho_{a|x}}) ~. \notag
\end{alignat}
The inequality holds because $\assemb{U\sigma_\lambda^*U^\dagger}$ is not the optimal solution of the SDP. Nevertheless, it is indeed a valid solution because it satisfies both constraints in \eq{\ref{eq:SDP}}:
\begin{alignat}{2}
    &\sum_\lambda D_\lambda(a|x)\sigma_\lambda' - \varrho_{a|x}' &&=  \sum_\lambda D_\lambda(a|x)U\sigma_\lambda^* U^\dagger - U\varrho_{a|x}U^\dagger \notag\\
    &~&&=U\left(\sum_\lambda D_\lambda(a|x)\sigma_\lambda^* - \varrho_{a|x}\right)U^\dagger \notag\\
    &~&&\geq 0 ~~ \forall ~ a, x~, \notag\\
    &\sigma_\lambda' = U\sigma_\lambda^*U^\dagger \geq 0  ~~ \forall ~ \lambda~. && \notag
\end{alignat}
Therefore, we arrive at another bound relation which is given as
\begin{equation}\label{eq:stsr_upper}
    \mathcal{STSR}(\assemb{\varrho_{a|x}'}) \leq \mathcal{STSR}(\assemb{\varrho_{a|x}}).
\end{equation}
There are some similar properties of \eq{\ref{eq:stsr_upper}} that have been discussed in Ref.~\cite{Rodrigo2015}.

By combining \eq{\ref{eq:stsr_lower}} and \eq{\ref{eq:stsr_upper}}, we find that $\mathcal{STSR}$ of the assemblage is invariant under unitary transformation:
\begin{equation}\label{eq:stsr_invariant}
    \mathcal{STSR}(\assemb{U\varrho_{a|x}U^\dagger}) = \mathcal{STSR}(\assemb{\varrho_{a|x}}),
\end{equation}
thus, we have completed the proof that perfect state transfer implies invariance of $\mathcal{STSR}$. $\blacksquare$

\section{Proof of \eq{\ref{eq:stsr_geq_D}} in the main text}\label{sec:proof_quantify_NSIT}

We now briefly summarize how to obtain the bound relation in \eq{\ref{eq:stsr_geq_D}} by additionally introducing two optimization problems ($\mathcal{R}_1$ and $\mathcal{R}_2$). We then show the bound relations of each optimization problems, namely (1) $\mathcal{STSR} \geq \mathcal{R}_1$, (2) $\mathcal{R}_1 \geq \mathcal{R}_2$, (3) $\mathcal{R}_2 \geq \mathcal{D}$. Due to the transitivity, we can complete the proof. % including the $\mathcal{STSR}$ in \eq{\ref{eq:robustness}} and the trace distance for estimating signaling effect ($\mathcal{D}$). Since all the optimization problems are real, we can complete the proof by transitive relation, namely $\mathcal{STSR} \geq \mathcal{R}_1 \geq \mathcal{R}_2 \geq \mathcal{D}$.

Once the NSIT condition is not satisfied, the marginal of the assemblage can be defined as $\varrho_x=\sum_a \varrho_{a|x}$. Motivated by the definition of $\mathcal{STSR}$ in \eq{\ref{eq:robustness}}, it is convenient to introduce the first optimization problem, namely
\begin{alignat}{1}\label{eq:R1}
    \mathcal{R}_1 (\assemb{\varrho_{a|x}}) &= \displaystyle \min_{r_1, \assemb{\tau_{x}}, \sigma} ~~r_1, \notag\\
    &\text{s.t.} ~~~\frac{\varrho_{x}+r_1\tau_x}{1+r_1}=\sigma ~~~\forall~~~ x,
\end{alignat}
where $\sigma$ and $\tau_{x}$ are arbitrary quantum states. It is easy to see that the above optimization problem is merely the robustness without the NSIT condition, and the corresponding SDP can be easily derived. We note that whether the above robustness has the corresponding resource theory is still vague~\cite{Chitambar2019}. It is easy to discover that the optimal solution for \eq{\ref{eq:robustness}} (denoted as $r^*$, $\assemb{\tau_{a|x}^*}$, and $\assemb{(\varrho_{a|x}^{\text{LHS}})^*}$) is a valid solution for \eq{\ref{eq:R1}} because it satisfies all the constraints in \eq{\ref{eq:R1}} by introducing $\tau_x=\sum_a \tau_{a|x}^*$ and $\sigma=\sum_a (\varrho_{a|x}^{\text{LHS}})^*$. Nevertheless, it may not be the optimal solution for \eq{\ref{eq:R1}}. Therefore, we have the first bound relation
\begin{equation}\label{eq:R1_upper}
    \mathcal{STSR}(\assemb{\varrho_{a|x}}) \geq \mathcal{R}_1(\assemb{\varrho_{a|x}}).
\end{equation}

Since the right-hand side of the constraint in \eq{\ref{eq:R1}} is independent of $x$, we can reformulate \eq{\ref{eq:R1}} as
\begin{alignat}{1}\label{eq:R1-2}
    \mathcal{R}_1 &(\assemb{\varrho_{a|x}}) = \displaystyle \min_{r_1, \assemb{\tau_{x}}} ~~r_1, \notag\\
    &\text{s.t.} ~~~ r_1(\tau_x - \tau_{x'})=\varrho_{x'}-\varrho_{x}~~~\forall~ x\neq x'.
\end{alignat}
With the above results, we can further introduce another optimization problem, namely
\begin{alignat}{1}\label{eq:R2}
    \mathcal{R}_2 &(\assemb{\varrho_{a|x}}) = \displaystyle \min_{r_2, \assemb{\tau_{x}}} ~~r_2, \notag\\
    &\text{s.t.} ~~~ r_2 ||\tau_x - \tau_{x'}||_1 = ||\varrho_{x}-\varrho_{x'}||_1~~~\forall~ x\neq x'.
\end{alignat}
The \eq{\ref{eq:R1-2}} and \eq{\ref{eq:R2}} satisfy the following bound relation
\begin{equation}\label{eq:R2_upper}
    \mathcal{R}_1(\assemb{\varrho_{a|x}}) \geq \mathcal{R}_2(\assemb{\varrho_{a|x}}).
\end{equation}
The inequality holds because the optimal solution in \eq{\ref{eq:R1-2}} is also a valid but not optimal solution for \eq{\ref{eq:R2}}.

Now, consider $r^*_2$ and $\assemb{\tau_x^*}$ to be the optimal solution for \eq{\ref{eq:R2}}, it must satisfy the constraint, namely
\begin{equation}
    r_2^* ~||\tau^*_x - \tau^*_{x'}||_1 = ||\varrho_{x}-\varrho_{x'}||_1~~~\forall~ x\neq x'. \notag
\end{equation}
Because the maximum value of the trace norm between two arbitrary quantum states is $2$, i.e., $||\tau_x^* - \tau^*_{x'}||_1 \leq 2$, we have
\begin{equation}
    r_2^* \geq \frac{1}{2}||\varrho_{x}-\varrho_{x'}||_1~~~\forall~ x\neq x', \notag
\end{equation}
or alternatively, 
\begin{equation}
    r^*_2 \geq \max_x ~\frac{1}{2}||\varrho_{x}- \varrho_{x'}||_1 ~~\forall~~x\neq x'. \notag
\end{equation}
Therefore, we arrive at the last bound relation
\begin{alignat}{1}\label{eq:R2_lower}
    \mathcal{R}_2(\assemb{\varrho_{a|x}}) &= r^*_2 \notag\\
    &\geq \max_x ~\frac{1}{2}\left|\left|\sum_a\varrho_{a|x}-\sum_a \varrho_{a|x'}\right|\right|_1 ~~\forall~~x\neq x' \notag\\
    &= \mathcal{D}(\assemb{\varrho_{a|x}}).
\end{alignat}
Finally, due to the transitivity, we complete the proof, namely
\begin{equation}
    \mathcal{STSR} \geq \mathcal{R}_1 \geq \mathcal{R}_2 \geq \mathcal{D}.~\blacksquare
\end{equation}

\section{The experimental results from different IBMQ devices}\label{App_diff_IBM_chips}

In \tab{\ref{tab:ibmq result compare}}, we show $\mathcal{STSR}$ under the process of the perfect QST with different IBMQ devices: 20-qubits almaden, 20-qubits boeblingen, 28-qubits cambridge, 5-qubits london, and 27-qubits paris. The circuit implementations for all are the same as the one introduced in the main text (see \sect{\ref{sec:state preparation}} and \sect{\ref{sec:quantum state transfer process}}). One can see that the different chips shows different performances of the QST. We thus can benchmark each chips under the QST tasks.

\begin{table}[!htbp]
    \caption{
        Experimental results of the perfect QST ($\theta=\pi$) from different IBMQ devices.
    }
    \label{tab:ibmq result compare}
    \begin{tabular}{ clccc }
        \hline
        \hline
        Devices & Transference routes & $~n~$ & $~\mathcal{STSR}~$ & Signaling \\
        \hline
        \multirow[t]{4}{*}{\vtop{\hbox{\strut almaden}\hbox{\strut (Mar, 2020)}}} & \multirow[t]{4}{*}{$0 \to 1 \to 2 \to 3 \to 4$} & 2 & 0.169 & 0.026 \\
        & & 3 & 0.130 & 0.021 \\
        & & 4 & 0.086 & 0.019 \\
        & & 5 & 0.040 & 0.021 \\
        \hline
        \multirow[t]{4}{*}{\vtop{\hbox{\strut almaden}\hbox{\strut (Mar, 2020)}}} & \multirow[t]{4}{*}{$5 \to 6 \to 7 \to 8 \to 9$} & 2 & 0.133 & 0.025 \\
        & & 3 & 0.040 & 0.025 \\
        & & 4 & 0.016 & 0.015 \\
        & & 5 & 0.018 & 0.018 \\
        \hline
        \multirow[t]{4}{*}{\vtop{\hbox{\strut boeblingen}\hbox{\strut (Mar, 2020)}}} & \multirow[t]{4}{*}{$0 \to 1 \to 2 \to 3 \to 4$} & 2 & 0.216 & 0.015 \\
        & & 3 & 0.202 & 0.018 \\
        & & 4 & 0.173 & 0.019 \\
        & & 5 & 0.129 & 0.026 \\
        \hline
        \multirow[t]{4}{*}{\vtop{\hbox{\strut boeblingen}\hbox{\strut (Jan, 2020)}}} & \multirow[t]{4}{*}{$0 \to 1 \to 2 \to 3 \to 4$} & 2 & 0.231 & 0.184 \\
        & & 3 & 0.198 & 0.153 \\
        & & 4 & 0.170 & 0.140 \\
        & & 5 & 0.140 & 0.138 \\
        \hline
        \multirow[t]{4}{*}{\vtop{\hbox{\strut boeblingen}\hbox{\strut (Mar, 2020)}}} & \multirow[t]{4}{*}{$5 \to 6 \to 7 \to 8 \to 9$} & 2 & 0.059 & 0.024 \\
        & & 3 & 0.133 & 0.025 \\
        & & 4 & 0.116 & 0.030 \\
        & & 5 & 0.033 & 0.030 \\
        \hline
        %\multirow[t]{4}{*}{\vtop{\hbox{\strut boeblingen}\hbox{\strut (Mar, 2020)}}} & \multirow[t]{4}{*}{$10 \to 11 \to 12 \to 13 \to 14$} & 2 & 0.1742 & 0.0287 \\
        %& & 3 & 0.0928 & 0.0307 \\
        %& & 4 & 0.1050 & 0.0242 \\
        %& & 5 & 0.0791 & 0.0247 \\
        %\hline
        \multirow[t]{4}{*}{\vtop{\hbox{\strut boeblingen}\hbox{\strut (Mar, 2020)}}} & \multirow[t]{4}{*}{$15 \to 16 \to 17 \to 18 \to 19$} & 2 & 0.025 & 0.021 \\
        & & 3 & 0.032 & 0.027 \\
        & & 4 & 0.010 & 0.010 \\
        & & 5 & 0.005 & 0.005 \\
        \hline
        \multirow[t]{4}{*}{\vtop{\hbox{\strut cambridge}\hbox{\strut (Jul, 2020)}}} & \multirow[t]{4}{*}{$0 \to 1 \to 2 \to 3 \to 4$} & 2 & 0.017 & 0.017 \\
        & & 3 & 0.006 & 0.006 \\
        & & 4 & 0.009 & 0.008 \\
        & & 5 & 0.011 & 0.011 \\
        \hline
        \multirow[t]{3}{*}{\vtop{\hbox{\strut london}\hbox{\strut (Oct, 2019)}}} & \multirow[t]{3}{*}{$0 \to 1 \to 3 \to 4$} & 2 & 0.203 & 0.022 \\
        & & 3 & 0.190 & 0.027 \\
        & & 4 & 0.154 & 0.029 \\
        \hline
        %\multirow[t]{3}{*}{\vtop{\hbox{\strut london}\hbox{\strut (Nov, 2019)}}} & \multirow[t]{3}{*}{$2 \to 1 \to 3 \to 4$} & 2 & 0.1948 & 0.0364 \\
        %& & 3 & 0.1884 & 0.0367 \\
        %& & 4 & 0.1381 & 0.0402 \\
        %\hline
        \multirow[t]{4}{*}{\vtop{\hbox{\strut paris}\hbox{\strut (Jul, 2020)}}} & \multirow[t]{4}{*}{$0 \to 1 \to 2 \to 3 \to 5$} & 2 & 0.208 & 0.074 \\
        & & 3 & 0.197 & 0.087 \\
        & & 4 & 0.148 & 0.039 \\
        & & 5 & 0.085 & 0.061 \\
        \hline
        \hline
    \end{tabular}
\end{table}

%\bibliography{stsr_benchmark_ref.bib}

%apsrev4-2.bst 2019-01-14 (MD) hand-edited version of apsrev4-1.bst
%Control: key (0)
%Control: author (8) initials jnrlst
%Control: editor formatted (1) identically to author
%Control: production of article title (0) allowed
%Control: page (0) single
%Control: year (1) truncated
%Control: production of eprint (0) enabled
%

\end{document}